\documentstyle[preprint,aps,epsfig,psfrag]{revtex}
\def\beq{\begin{equation}}
\def\eeq{\end{equation}}
\def\barr#1{\begin{array}{#1}}
\def\earr{\end{array}}
\def\beqar{\begin{eqnarray}}
\def\eeqar{\end{eqnarray}}
\def\beqars{\begin{eqnarray*}}
\def\eeqars{\end{eqnarray*}}
\def\plb#1#2#3#4{#1, Phys. Lett. {\bf B#2} (#3) #4}
\def\npb#1#2#3#4{#1, Nucl. Phys. {\bf B#2} (#3) #4}
\def\prd#1#2#3#4{#1, Phys. Rev. {\bf D#2} (#3) #4}
\def\prl#1#2#3#4{#1, Phys. Rev. Lett. {\bf #2} (#3) #4}
\def\prog#1#2#3#4{#1, Prog. Theor. Phys. {\bf #2} (#3) #4}

\def\hepph#1#2#3{#1, #2, hep-ph/#3}

\begin{document}
\draft
\preprint{\begin{tabular}{l}
\hbox to\hsize{August, 1998 \hfill KAIST-16/98}\\[-3mm]
\hbox to\hsize{hep-ph/9808478 \hfill }\\[5mm] \end{tabular} }
\vspace{1cm}
\title{ Extraction method of $\gamma$ from semi-inclusive
$b \rightarrow D s$ decays}
 
\vspace{2cm}
   
\author{ 
Ji-Ho Jang \footnote{e-mail~:~jhjang@chep6.kaist.ac.kr} 
}
\address{
Department of Physics, Korea Advanced Institute of Science and Technology,\\
Taejon 305-701, Korea}
\maketitle
            
\begin{abstract}
We propose a new method to extract a $CP$ angle $\gamma$ from
semi-inclusive two-body nonleptonic decays, 
$b \rightarrow D^0(\bar{D}^0) s$ and $b \rightarrow D_{CP} s$.
This method is free from the unknown long distance strong interaction
effects and gives theoretically cleaner signal than the similar method
using exclusive nonleptonic decays, $B \rightarrow D K$.
We can determine $\gamma$ with 4-$\sigma$ accuracy for
$40^o \lesssim \gamma \lesssim 160^o$.
\end{abstract}


\newpage

\narrowtext
\tighten

In the standard model ( SM ), the source of $CP$ asymmetry is 
one complex phase in 
the Cabibbo-Kobayashi-Maskawa ( $CKM$ ) matrix element \cite{CKM}.
Until now, only one experimental evidence of $CP$ violation is found in
$K_L \rightarrow \pi \pi$ decay and it comes 
dominantly from $K^0-\bar{K}^0$ mixing.  Hence it should be important 
to probe the $CP$ violation in the $B$ systems 
in the future experiment in order to test the SM scenario of $CP$ violation.
One of the important objects to investigate
$CP$ violation is the unitary triangle
which includes three angles $\alpha, \beta$ and $\gamma$.
The angles $\beta$ and $\gamma$ are related to $V_{td}$ and $V_{ub}$ 
respectively and the angle $\alpha$ is obtained using the unitary  relation 
$\alpha + \beta + \gamma = \pi$. 
The recent numerical constraints of the three angles are given \cite{aliburas} :
\beqar
\label{exp1}
-1.0 \le &\sin 2\alpha& ~\le 1.0,
\\ \nonumber
0.30 \le &\sin 2\beta& ~\le 0.88,
\\ \nonumber
0.27 \le &\sin^2 \gamma& ~\le 1.0.
\eeqar

There are many suggestions to determine independently the angles of the
unitary triangle.
For example, the angle $\alpha$ can be  determined by 
$B \rightarrow \pi \pi$ modes if their gluonic penguin
pollutions can be removed using the isospin relation \cite{GL90}.
The $B \rightarrow J/\psi K_S$ decay is gold-plated mode 
to determine the angle $\beta$ because the $CKM$ angles 
from decay processes are almost canceled in the rate asymmetry
and it can be unambiguously determined
by $B^0 - \bar{B}^0$ mixing which is related to the angle $\beta$.
The angle $\gamma$ may be obtained by 
$B \rightarrow D K$ modes \cite{GLW91,Dunietz91}.
But it is noted in \cite{ADS97} that this method has the 
experimental difficulties 
because the final $\bar{D}^0$ meson should be identified using
$\bar{D}^0 \rightarrow K^+ \pi^-$, but it is difficult to distinguish it
from doubly Cabibbo suppressed  $D^0 \rightarrow K^+ \pi^-$ following color
and $CKM$ allowed $B^- \rightarrow D^0 K^-$. There are some variant methods
to overcome these difficulties \cite{ADS97,Gronau98,GR98,JK98}. 
In Ref.\cite{ADS97}, the interference between 
$B^- \rightarrow D^0 [\rightarrow \pi^- K^+] K^-$ decay
and $B^- \rightarrow \bar{D}^0 [\rightarrow \pi^- K^+] K^-$ decay
is used. The extraction method of the angle $\gamma$  using the 
color-allowed decays only is proposed in Ref.\cite{Gronau98}.
In Ref. \cite{GR98,JK98}, authors proposed the extraction method of
$\gamma$ using the isospin relation and neglecting the annihilation
diagrams.

Other methods to constrain the angle $\gamma$ from $B \rightarrow K \pi$ modes
has been also proposed in Ref.\cite{FM98}. 
However the long distance strong interaction effects
might destroy the validity of this method \cite{rescattering}. 
These uncertainties and electro-weak penguin pollutions 
can be removed by using the 
$B \rightarrow K K$ modes and $SU(3)$ flavor symmetry \cite{Fleischer98}.
Such rescattering effects may be 
potentially important in any exclusive decays of $B$ meson decays
and it is important to remove hadronic uncertainties coming
from the rescattering effects.

It is noted by several authors that inclusive and semi-inclusive decays 
of $B$ meson could show large $CP$ violation \cite{inclusive,insemi,BDHP98}.
In Ref.~\cite{inclusive}, the authors estimate the rate asymmetry using the
absorptive part of the decay amplitude. Gerard and Hou \cite{inclusive}
noted that $CPT$ theorem is violated if one does not include all diagrams
of the same order.
The $CP$ violation in the semi-inclusive charmless, single charm and 
double charm transition is considered in Ref.~\cite{insemi}. 
Moreover as the large cancellation between the semi-inclusive decay rates, 
the $CP$ violation effects in the totally inclusive decays are 
expected to be tiny.  However one can expect the large $CP$ violation in
the each of the semi-inclusive decays. 
Authors in Ref.~\cite{BDHP98} investigate the $CP$ violation in the
quasi-inclusive decays of the type $B \rightarrow K^{(*)} X$ 
that the strange quark is only included in $K^{(*)}$ meson.
The interference between the tree level process 
$b \rightarrow u \bar{u} s$ and the one loop process
$b \rightarrow s g^* \rightarrow s q \bar{q}$ gives the direct 
$CP$ violation. 

In the Ref.\cite{AS98}, authors proposed a systematic method of experimental
search for two-body hadronic decays of the $b$-quark of the type
$b \rightarrow quark(q) + meson(M)$. They have the well-defined experimental
signature because the spectrum of the meson energy should be 
a peak centered around $( m_b^2 + m_M^2 - m_q^2 )/2 m_b$ with
a spread of a few hundred MeV. The energy of outgoing quark
will be similar to that of the meson and become $\sim 2$ GeV numerically.
The hadronization process of the quark will lead to low average
multiplicity about 3/event. They insisted that the combinatorics problem
in discriminating against background is not so difficult.
They considered the tree$\times$penguin and penguin$\times$penguin type
processes to estimate the partial rate asymmetries and 
to consider electroweak penguin dominated branching ratios. 

In this letter, we suggest a new extraction method of $\gamma$ using
the semi-inclusive two-body decays of the type
$b \rightarrow D^0(\bar{D}^0) s$ and $b \rightarrow D_{CP} s$
which come form only tree diagrams ( see Fig.~1 ). 
These modes have the same advantage as other semi-inclusive decays 
given in Ref.~\cite{AS98}.
The theoretical values of the decay
rates would be  less uncertain than the exclusive modes because the 
hadronic form factors are replaced with the calculation of quark diagrams.
As these decay modes are semi-inclusive processes and final states involve
a kind of isospin state, we also expect that they are
free from the long distance effects of strong interaction.
Because we consider decays of $b-$quark, all kinds of $B$ hadrons can be used
in this analysis.

The effective Hamiltonian relevant to $b \rightarrow D s$ decays
is given by
\beqar
{\cal H}_{eff} &=& -\frac{G_F}{\sqrt{2}}
 \left[ V_{cb} V^*_{us} \left\{ c_1 (\bar{s}u)_{V-A} (\bar{c}b)_{V-A}
                              + c_2 (\bar{c}u)_{V-A} (\bar{s}b)_{V-A}
     \right\} \right. \\ \nonumber
&&~~~~
   + \left. V_{ub} V^*_{cs} \left\{ c_1 (\bar{s}c)_{V-A} (\bar{u}b)_{V-A}
                              + c_2 (\bar{u}c)_{V-A} (\bar{s}b)_{V-A}
     \right\} + h. c. \right],
\eeqar
where $c_{1(2)}$ are the Wilson coefficients and 
the subindex $V-A$ denotes
$\gamma_{\mu} ( 1 - \gamma_5 )$ structure.

Let us introduce the relevant amplitudes as follows,
\beqar
\label{amplitude}
A(b \rightarrow D^0 s) =& A(\bar{b} \rightarrow \bar{D}^0 \bar{s})
&~= {\cal A}, \\ \nonumber
e^{i \gamma} A(b \rightarrow \bar{D}^0 s) =&
e^{- i \gamma} A(\bar{b} \rightarrow D^0 \bar{s}) &~= 
{\cal A} ~R_b,
\eeqar 
where $R_b = \sqrt{\rho^2 + \eta^2} = 0.36 \pm 0.08$.  
The initial state $b (\bar{b})$ is isosiglet and the final states 
are isodoublet. 
Hence these processes are described
by the effective Hamiltonian of $|\Delta I| = 1/2$ 
and all amplitudes are given by the  a complex number
${\cal A}$ with same strong phases and different $CKM$ factors.
It is a main advantage of using these semi-inclusive decays that
there is no relative strong phase in the above amplitudes.

The semi-inclusive decay rates into flavor specific states of 
$D$ meson are given by
\beqar
\label{flavorrate}
\Gamma(b \rightarrow D^0 s) =& \Gamma(\bar{b} \rightarrow \bar{D}^0 \bar{s} )&~
= \left|{\cal A}\right|^2 ~F_{P.S},
\\ \nonumber
\Gamma(b \rightarrow \bar{D}^0 s) =& 
\Gamma(\bar{b} \rightarrow D^0 \bar{s} )&~
= \left|{\cal A}\right|^2 ~R_b^2 ~F_{P.S},
\eeqar
where $F_{P.S.}$ is phase space factor and we neglect the small phase
space differences.

On the other hand, using the definition of $CP$ eigenstates of $D$ mesons,
$D_{1(2)}=\frac{1}{\sqrt{2}} ( D^0 \pm \bar{D}^0 )$, and neglecting
the small $D^0 - \bar{D}^0$ mixing, we obtain the decay rates into 
$CP$ eigenstate of final $D$ meson: 
\beqar
\label{cpeigenrate}
\Gamma(b \rightarrow D_1 s) =& \Gamma(\bar{b} \rightarrow
D_1 \bar{s})&~=
\frac{1}{2} ~\left|{\cal A}\right|^2 ~F_{P.S.} 
~( 1 + R_b^2 + 2 R_b \cos \gamma ), 
\\ \nonumber
\Gamma(b \rightarrow D_2 s) =& \Gamma(\bar{b} \rightarrow
D_2 \bar{s})&~=
\frac{1}{2} ~\left|{\cal A}\right|^2 ~F_{P.S.} 
~( 1 + R_b^2 - 2 R_b \cos \gamma ). 
\eeqar

Decay rate ratios ${\cal A}_i$ between $CP$ eigenstates and 
flavor specific states in the final $D$ mesons are defined as follows,
\beqar
\label{definitionAi}
{\cal A}_i &=& 
\frac{\Gamma(b \rightarrow D_i s) + \Gamma(\bar{b} \rightarrow D_i \bar{s})}
{\Gamma(b \rightarrow D_1 s)+\Gamma(\bar{b} \rightarrow D_1 \bar{s})+
\Gamma(b \rightarrow D_2 s)+\Gamma(\bar{b} \rightarrow D_2 \bar{s})}
\\ \nonumber
&=&
\frac{\Gamma(b \rightarrow D_i s) + \Gamma(\bar{b} \rightarrow D_i \bar{s})}
{\Gamma(b \rightarrow D^0 s)+\Gamma(\bar{b} \rightarrow \bar{D}^0 \bar{s})+
\Gamma(b \rightarrow \bar{D}^0 s)+\Gamma(\bar{b} \rightarrow D^0 \bar{s})},
\eeqar
where $i = 1,2$ is $CP$-even and odd eigenstates respectively and 
in the second line, we use the following relation coming
from Eq.~(\ref{flavorrate}) and (\ref{cpeigenrate}) :
\beqar
&&\Gamma(b \rightarrow D^0 s)+\Gamma(\bar{b} \rightarrow \bar{D}^0 \bar{s})+
  \Gamma(b \rightarrow \bar{D}^0 s)+\Gamma(\bar{b} \rightarrow D^0 \bar{s})
\\ \nonumber
&=&\Gamma(b \rightarrow D_1 s) +\Gamma(\bar{b} \rightarrow D_1 \bar{s}) +
  \Gamma(b \rightarrow D_2 s) +\Gamma(\bar{b} \rightarrow D_2 \bar{s}).
\eeqar

Using Eq.(\ref{flavorrate}), we can simplify the ratio 
${\cal A}_i$ as follows,
\beq
\label{resultAi}
{\cal A}_i = \frac{1}{2} \pm \frac{R_b}{1+R_b^2} \cos \gamma,
\eeq
where $\pm$ correspond to $i = 1,2$ respectively.
Note that these simple relations come from the fact that
there is no relative strong phase between
$A( b \rightarrow D^0 s)$ and $A( b \rightarrow \bar{D}^0 s)$
in Eq.~(\ref{amplitude}).

We can also define a combined asymmetry ${\cal R}$ 
between $CP$-even and odd states as follows,
\beqar
\label{definitionR}
{\cal R} &=& {\cal A}_1 - {\cal A}_2
\\ \nonumber
&=&
\frac{ \{ \Gamma(b \rightarrow D_1 s) + 
              \Gamma(\bar{b} \rightarrow D_1 \bar{s})
       \} -
       \{ \Gamma(b \rightarrow D_2 s) + 
              \Gamma(\bar{b} \rightarrow D_2 \bar{s})
       \} }
     { \{ \Gamma(b \rightarrow D_1 s) + 
             \Gamma(\bar{b} \rightarrow D_1 \bar{s})
       \}
        +
       \{ \Gamma(b \rightarrow D_2 s) + 
             \Gamma(\bar{b} \rightarrow D_2 \bar{s})
       \} },
\eeqar

Using Eq.(\ref{cpeigenrate}) or Eq(\ref{resultAi}),
$\cos \gamma$ is related to the asymmetry ${\cal R}$ as
\beq
\label{resultR}
\cos \gamma = \frac{{\cal R} (1 + R_b^2)}{2 R_b}.
\eeq
The sum of initial $b$ and $\bar{b}$ states
is used in the definition of 
the ratio ${\cal A}_i$ in Eq.~(\ref{definitionAi}) and the asymmetry
${\cal R}$ in Eq.~(\ref{definitionR}).
Hence there is no need tagging of the initial $b$ and $\bar{b}$ states
and it is an experimental advantage of this method.

From Eq.~(\ref{resultAi}),  (\ref{resultR}) and using the present
experimental value of $R_b = 0.36 \pm 0.08$, we can obtain the bound
of the ratios ${\cal A}_i$ and the asymmetry ${\cal R}$ :
\beqar
0.18 \leq &{\cal A}_i& \leq 0.82,
\\ \nonumber
-0.64 \leq &{\cal R}& \leq 0.64.
\eeqar
If the ratios ${\cal A}_i$ and ${\cal R}$ are determined
between the above range in the future experiments, 
we can constrain the angle $\gamma$ using the experimental data.

Let's probe the feasibility of this method in determining the angle $\gamma$.
We need to know the branching ratio of each mode and the detection
efficiences.
In order to estimate the branching ratio of the relevant decay,
we use the factorization approximation and obtain the
relevant transition amplitudes :
\beqar
A(b \rightarrow D^0 s) &=&
\frac{G_F}{\sqrt{2}} ~V_{cb} V^*_{us} ~a_2
< D^0 | \bar{c}u_- |  0 >< s | \bar{s}b_- | b >, \\ \nonumber
A(b \rightarrow \bar{D}^0 s) &=&
\frac{G_F}{\sqrt{2}} ~V_{ub} V^*_{cs} ~a_2
< \bar{D}^0 | \bar{u}c_- |  0 >< s | \bar{s}b_- | b >, 
\eeqar
where $a_2 = c_2 + c_1/N_c$ and 
$\bar{q} q'_- = \bar{q} \gamma_{\mu} ( 1 - \gamma_5 ) q'$.

The branching ratio is given by
\beqar
&&Br(b \rightarrow D^0 s) =
\left| \frac{V_{cb} V^*_{us}}{V_{ub} V^*_{cs}} \right|^2
Br(b \rightarrow \bar{D}^0 s) 
\\ \nonumber
&& = \frac{G_F^2}{8 \pi m_b} ~| V_{cb} V^*_{us}|^2 ~a_2^2 ~f_D^2 ~R
 ~\tau_b
 ~[ 2(p_s \cdot p_D)(p_b \cdot p_D) - (p_s \cdot p_b) m_D^2 ]
\\ \nonumber
&& \simeq 0.67 \times 10^{-4},
\eeqar
where $R = (1+R_1^2+R_2^2-2 R_1 R_2 - 2 R_1 - 2 R_2 )^{1/2}$
with $R_1 = (m_s/m_b)^2$ and $R_2 = ( m_D /m_b)^2$.
In this numerical calculation, we use the following parameter set : 
$\tau_b = 1.55 \times 10^{-12}$sec, 
$a_2 = 0.21, f_D = 200 ~MeV, m_s = 120 ~MeV,
m_D = 1.87 ~GeV$ and $m_b = 4.8 ~GeV$.
The values of the branching ratios depend on the specific parameter set, 
the factorization assumption and the Fermi motion of the $b$-quark
in the $B$ mesons. However the above typical values
are enough to estimate the errors in determining the $\gamma$.

In order to estimate the uncertainty in the determination of the angle
$\gamma$, we assume $3 \times 10^8$ $B\bar{B}$ events in $B$ factories
using $e^+ e^-$ annihilation at the $\Upsilon (4S)$ resonance.
Tagging of $D^0(\bar{D}^0)$ is used 
$D^0 \rightarrow K^- \pi^+, D^0 \rightarrow K^- \pi^+ \pi^+ \pi^-$ and
$D^0 \rightarrow K^- \pi^+ \pi^0$ modes 
and its total efficiency is about $0.25$.
The $CP$ even-state $D_1$ is identified by 
$D_1 \rightarrow \pi^+ \pi^-, K^+ K^-$ and the observation rate is
$5 \times 10^{-2}$ which is quoted in Ref. \cite{Dunietz91}.
The tagging of $CP$ odd state $D_2$
uses the $D_2 \rightarrow K_s \pi^+ \pi^-$
whose branching ratio is $5.4 \%$. The observation rate is $3.6 \%$ as
$K_S$ is identified by $K_S \rightarrow \pi^+ \pi^-$ mode whose branching
fraction is about 2/3.

The number of the observable events can be obtained by the product of
the number of the $B \bar{B}$ event, the branching ratio of the each modes
and the detection efficiencies of the final particles. 
The statistical error in the branching ratio is approximately given by
$\Delta Br/Br \approx 1/\sqrt{N_{\rm obs}}$.
The numerical values of the statistical error become about 
$1.4 \%$ and $3.9 \%$ for $b(\bar{b}) \rightarrow D^0 s (\bar{D}^0 \bar{s})$
and $b(\bar{b}) \rightarrow \bar{D}^0 s (D^0 \bar{s})$, respectively.
For the modes with $CP$ eigen states in the final states, the values depend
on the angle $\gamma$. Presenting the values in order of 
$b \rightarrow D_1 s ( \bar{b} \rightarrow D_1 \bar{s})$ and 
$b \rightarrow D_2 s ( \bar{b} \rightarrow D_2 \bar{s})$, they become
$3.4 \% (4.0 \%)$ and $6.3 \% (7.4 \%)$ for $\gamma = 30^o$,
$4.0 \% (4.7 \%)$ and $4.4 \% (5.2 \%)$ for $\gamma = 80^o$ and
$5.9 \% (6.9 \%)$ and $3.4 \% (4.0 \%)$ for $\gamma = 140^o$, respectively.

Using this information, we can estimate the error in determining $\gamma$.
The result is given in Fig.~2, where the horizontal and vertical axis represent
$\gamma$ and $\Delta \gamma$ in degrees, respectively.
The real line and dashed line present the error in $\gamma$ determined using
the ratio ${\cal A}_1$ and ${\cal A}_2$.
The dotted line is error plot using the combined asymmetry ${\cal R}$.
They all give the similar results.
From $40^o$ to $160^o$, we can determine $\gamma$ with $4$-$\sigma$ accuracy.
We also investigate the possibility to extarct the information of 
the angle $\gamma$ in the smaller number of $B \bar{B}$ events, $3 \times 10^7$.
The result is given in Fig.~3. 
Even in this case, our method may give the reliable 
results that the angle can be determined with 2-$\sigma$ accuracy 
for $40^o \lesssim \gamma \lesssim 60^o$ and 3-$\sigma$ accuracy for 
$60^o \lesssim \gamma \lesssim 160^o$.

In conclusion, we proposed a new extraction method of $\gamma$ using the
semi-inclusive decays of the type $b \rightarrow D s, \bar{D} s, D_{CP} s$.
These decays are relevant to only tree diagram (Fig.~1) and 
there is no relative
strong phase depending on the final states because the final states 
are related to the semi-inclusive modes and only one kind of isospin state. 
Then we can simply relate $\cos \gamma$ to the $R_b$ and the ratios 
${\cal A}_i, {\cal R}$ which should be determined in future experiment.
Since we consider decays of $b-$quark, all kinds of $b-$ hadrons can be 
used in this analysis.
We also probe the feasibility of the suggested  method in this paper
by estimating the statistical errors in determination of  
the angle $\gamma$ which is given in Fig.~2. 
The value of $\gamma$ should be determined with
$4$-$\sigma$ accuracy for $40^o \lesssim \gamma \lesssim 160^o$
using $3 \times 10^8 B \bar{B}$ events in our method. 

\acknowledgements
I am grateful to Prof. Pyungwon Ko for helpful discussions, 
encouragement and reading the manuscript.


\psfrag{Xbb}[][][0.85]{$b$}
\psfrag{Xss}[][][0.85]{$s$}
\psfrag{Xc1}[][][0.85]{$c$}
\psfrag{Xc2}[][][0.85]{$\bar{c}$}
\psfrag{Xu1}[][][0.85]{$\bar{u}$}
\psfrag{Xu2}[][][0.85]{$u$}
\psfrag{Xp1}[][][0.85]{$D^0$}
\psfrag{Xp2}[][][0.85]{$\bar{D}^0$}
\psfrag{XV1}[][][0.85]{$V_{cb}$}
\psfrag{XV2}[][][0.85]{$V_{us}^*$}
\psfrag{XV3}[][][0.85]{$V_{ub}$}
\psfrag{XV4}[][][0.85]{$V_{cs}^*$}

\begin{figure}[h]
\centerline{\epsffile{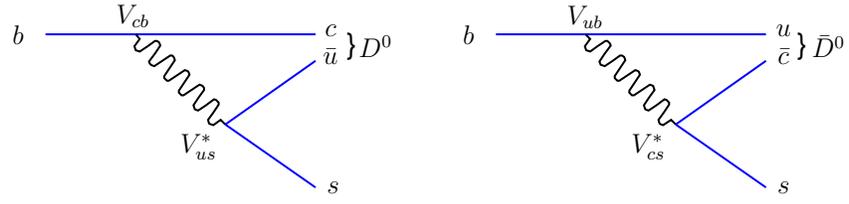}}
\caption{ Feynman diagram for $b \rightarrow D s$ decay modes}
\label{fig1}
\end{figure}

\begin{figure}[h]
\centerline{\epsffile{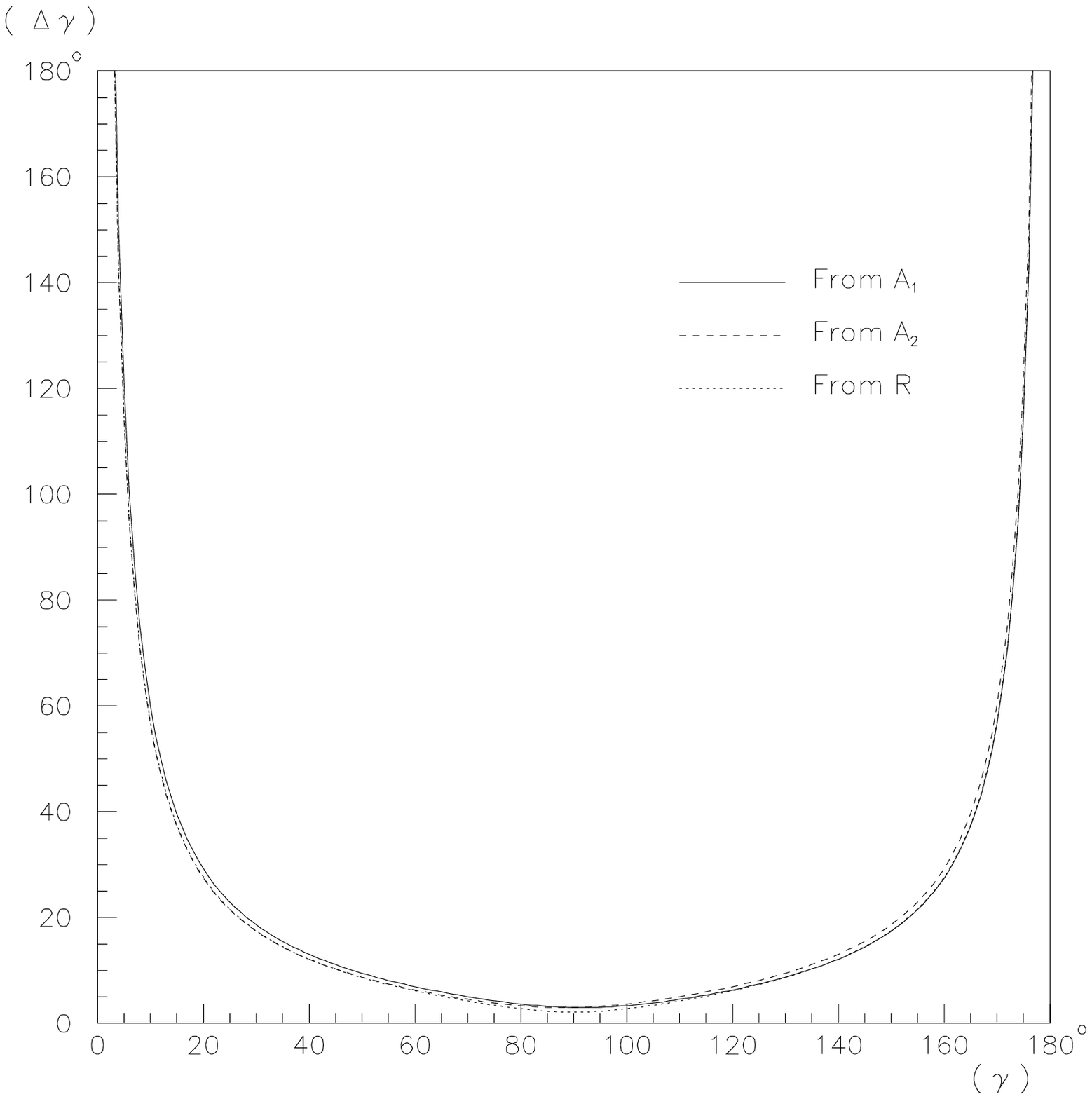}}
\caption{The $\Delta \gamma$ ( error ) plot in the determination of $\gamma$
assuming $3 \times 10^{8}$ $B$'s at $B$ factories :
real line is using $A_1$, dashed line is using $A_2$ and 
dotted line is using $R$
}
\label{fig2}
\end{figure}

\begin{figure}[h]
\centerline{\epsffile{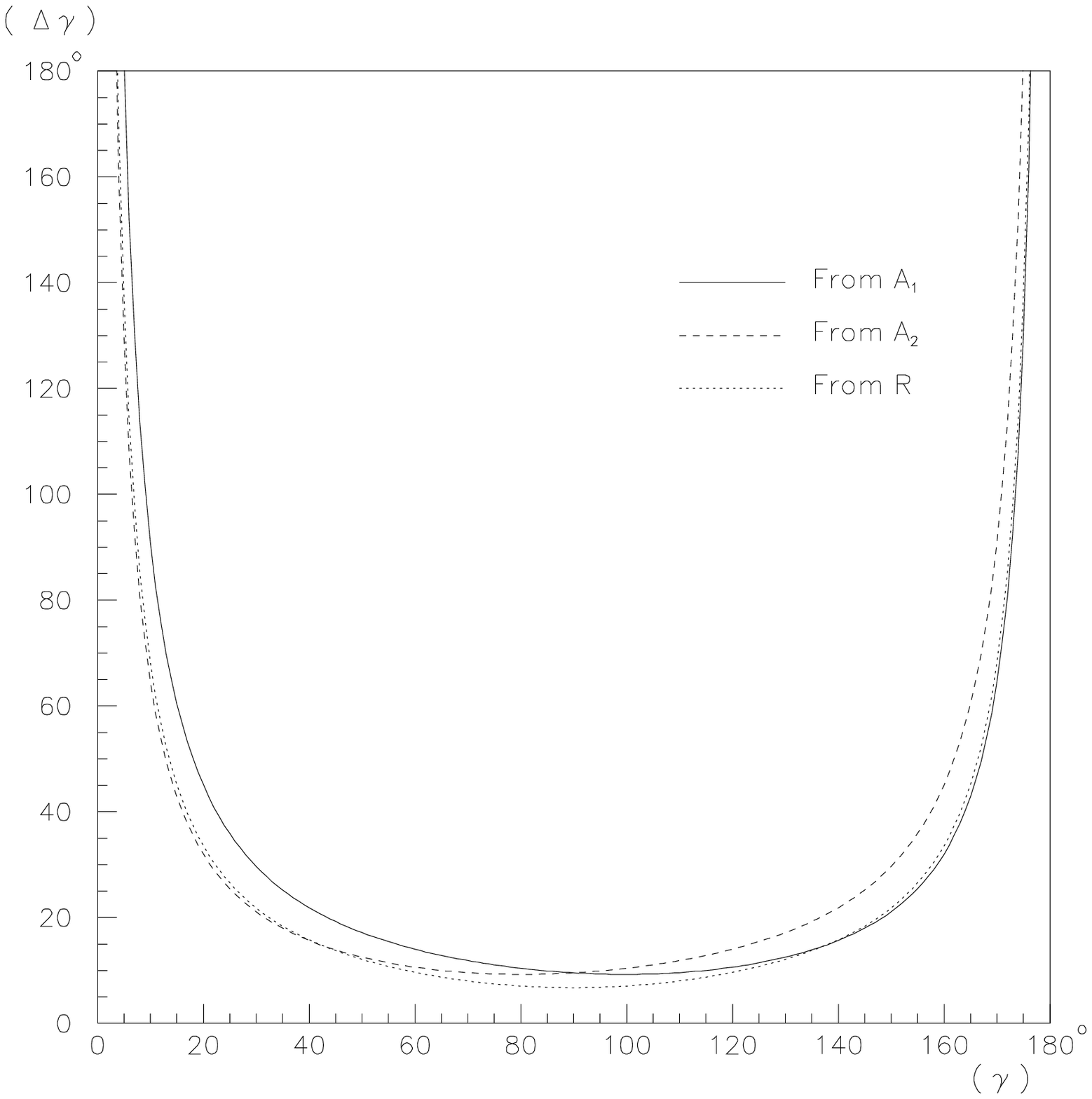}}
\caption{The $\Delta \gamma$ ( error ) plot in the determination of $\gamma$
assuming $3 \times 10^{7}$ $B$'s at $B$ factories :
real line is using $A_1$, dashed line is using $A_2$ and 
dotted line is using $R$
}
\label{fig2}
\end{figure}


\begin{thebibliography}{99}

\bibitem{CKM}
\prl{N. Cabibbo}{10}{1963}{531};
\prog{M. Kobayash and K. Maskawa}{49}{1973}{652}.


\bibitem{aliburas}
A. J. Buras, TUM-HEP-299-97, hep-ph/9711217 ;
J.L. Rosner, Nucl. Instrum. Meth. {\bf A 408}, 308 (1998) ;
A. Ali, hep-ph/9801270.

\bibitem{GL90}
\prl{M. Gronau and D. London}{65}{1990}{3381}.

\bibitem{GLW91}
\plb{M. Gronau and D. London}{253}{1991}{483};
\plb{M. Gronau and D. Wyler}{265}{1991}{172}.

\bibitem{Dunietz91}
\plb{I. Dunietz}{270}{1991}{75}.

\bibitem{ADS97}
\prl{D. Atwood, I. Dunietz and A. Soni}{28}{1997}{3257}.

\bibitem{Gronau98}
\hepph{M. Gronau}{CALT-68-2159}{9802315}.

\bibitem{GR98}
\hepph{M. Gronau and J. L. Rosner}{EFI-98-29, FERMILAB-PUB-98-227-T}{
9807447}.

\bibitem{JK98}
\hepph{J. H. Jang and P. Ko}{KAIST-14/98, SNUTP 98-079}{9807469}.

\bibitem{FM98}
\prd{R. Fleischer and T. Mannel}{57}{1998}{2752}.

\bibitem{rescattering}
\hepph{J. M. Gerard and J. Weyers}{UCL-IPT-97-18}{9711469};
\plb{M. Neubert}{424}{1998}{152};
\prd{A. F. Falk, A. L. Kagan, Y. Nir and A. A. Petrov}{57}{1998}{4290};
\hepph{D. Atwood and A. Soni}{AMES-HET-97-10}{9712287}.

\bibitem{Fleischer98}
\hepph{R. Fleischer}{CERN-TH/98-60}{9802433};
CERN-TH/98-128, hep-ph/9804319;
\hepph{M. Gronau and J. L. Rosner}{EFI-98-23}{9806348}.

\bibitem{AS98}
\hepph{D. Atwood and A. Soni}{AMES-HET-98-6,BNL-HET-98/19}{9805393}.

\bibitem{inclusive}
\prl{M. Bander, D. Siverman and A. Soni}{43}{1979}{242};
\prd{J. M. Gerard and W. S. Hou}{43}{1991}{2909};
\npb{H. Simma, G. Eilam and D. Wyler}{352}{1991}{367};
\prd{L. Wolfenstein}{43}{1991}{151};
\hepph{A. Lenz, U. Nierste and G. Ostermaier}{DESY-97-208}{9802202}.

\bibitem{insemi}
\hepph{I. Dunietz}{FERMILAB-PUB-97/323-T}{9806521};
\plb{M. Beneke, G. Buchalla and I. Dunietz}{393}{1997}{132};
J. Bernabeu and C. Jarlskog $ibid.$ {\bf B301} (1993) 273;
\prd{I. Dunietz and R. G. Sachs}{37}{1988}{3186};
(E) $ibid.$ {\bf D39} (1989) 3515.

\bibitem{BDHP98}
T. T. Browder, A. Datta, X. G. He and S. Pakvasa, hep-ph/9807280;
Phys. Rev. {\bf D57} (1998) 6829.
\end{thebibliography}
\end{document}